\documentclass[aps,%
floatfix,%
final,%
notitlepage,%
oneside,%
onecolumn,%
nobibnotes,%
nofootinbib,%
superscriptaddress,%
showpacs,%
prd,%
11pt
]%
{revtex4}

\usepackage{amsmath}
\usepackage{graphicx}
\usepackage{multirow}

\begin{document}

\large
\title{$\Upsilon$-meson pair production at LHC}

\author{\firstname{A.~V.}~\surname{Berezhnoy}}
\email{Alexander.Berezhnoy@cern.ch}
\affiliation{SINP of Moscow State University, Moscow, Russia}
\author{\firstname{A.~K.}~\surname{Likhoded}}
\email{Anatolii.Likhoded@ihep.ru}
\affiliation{Institute for High Energy Physics, Protvino, Russia}
\author{\firstname{A.~A.}~\surname{Novoselov}}
\email{Alexey.Novoselov@cern.ch}
\affiliation{Institute for High Energy Physics, Protvino, Russia}

\begin{abstract}
\normalsize
Theoretical  predictions for $pp\to2\Upsilon+X$ cross section at $\sqrt{s}=8$
TeV for the LHCb and ATLAS kinematical conditions are obtained.
A possibility to observe the new hypothetical particles  containing two 
valence $b$-quarks and two valence $\bar{b}$-quarks is discussed.
\end{abstract}

\pacs{13.85.Fb,  14.40.Rt}

\maketitle

\section{Introduction}

In the recent report of the LHCb collaboration \cite{Aaij:2011yc}  a measurement of the $J/\psi$ pair production at  $pp$ collision energy of 8~TeV  was 
presented. The experimental cross section value of $5.1 \pm 1.1~\mathrm{nb}$ appeared to be close 
to the predicted one in the article \cite{Berezhnoy:2011xy}, in which the
leading order (LO) QCD calculations in the color singlet (CS) scheme was performed. This calculations take into account only the single parton scattering reactions (SPS). 
However,  the double parton scattering (DPS) can also contribute to the  $J/\psi$ pair production at LHC due to the large gluon luminosity. The cross section value of double 
$J/\psi$ production was  roughly estimated  within DPS mechanism is close to the SPS prediction. 

The  calculation for the process $gg\to J/\psi J/\psi$ within LO QCD + CS + SPS is a well defined procedure  and give a reliable results. The calculation uncertainties are determined by a choice of scales,  parton density functions, and a mass of heavy quark. 
 The cross section value within this  approach can be estimated as $4 \pm 2~\mathrm{nb}$~\cite{Berezhnoy:2011xy}. 
Of cause there some open questions in this calculation: the contribution from the NLO diagrams, the influence  of the relative quark motion in the hard part of the matrix element, and the accounting of transverse momenta of initial gluons.  

Recently the estimation of some high order contributions to the  $J/\psi$ pair production has been done in~\cite{Baranov:2012re}.
The effects due to the transverse momenta of initial gluons was studied in~\cite{Baranov:2011zz}  within the $k_T$-factorization approach.
Also it has been shown that the quark internal motion could influence the cross section distribution shape~\cite{Berezhnoy:2011xy}, as well as the cross section value~\cite{Martynenko:2012tf}.
However, all these problems need further investigation.

It is worth to  note, that the standard mechanism (pQCD + CS + SPS) does not describe well enough
the invariant mass distribution of $J/\psi$ pair production \cite{Aaij:2011yc}, however it is shown that the quark motion in quarkonia could improve the agreement~\cite{Berezhnoy:2012xq}. 
The $p_T$ and $y$ distributions for the double $J/\psi$ production were not reported yet~\cite{Aaij:2011yc}. 

In the case of double $\Upsilon$ production quarks in quarkonia have a smaller 
velocity~\cite{Bodwin:1994jh}, thus the uncertainties caused by the  quark internal motion are smaller than for the case of 
double $J/\psi$ production. Therefore the formula for the cross section of the process $gg \to (Q\bar Q) + (Q\bar Q) $ obtained within leading order of pQCD~\cite{Humpert:1983yj} should describe the 
data on the double $\Upsilon$ production more precisely than the data on the double $J/\psi$ production.

What concerns a possible DPS contribution, only simplified models can be recently used to estimate its contribution due to the absence of the two-particle distribution functions for partons in a proton. 
This leads to the unknown uncertainties in the cross section value. Using the LHCb experimental cross section value for the inclusive single $J/\psi$ production~\cite{Aaij:2011jh}
and the  conventional value for the effective DPS cross section, the cross section value of $J/\psi$ pair production in the LHCb acceptance~\cite{Abe:1997xk,Abazov:2009gc} can be estimated as $~4$~nb.
As it was noted, this value is close to the result of calculation in the SPS approach.

Later the associated production of $J/\psi$ and open charm and the double open charm production was measured 
in the same experiment~\cite{LHCb-PAPER-2012-003}. It appeared that the standard mechanism (pQCD + SPS) underestimates the experimental data in this channels by a factor of $3\div 10$. The needed cross section values can be obtained 
within the DPS approach, which implies an independent production of two $c \bar{c}$-pairs. However it is worth to mention, 
that the DPS approach can not describe kinematical distribution shapes for the both these cases.

In this paper we show that for the double $\Upsilon$ production in the LHCb conditions the DPS contribution 
is negligibly small, and therefore the double $\Upsilon$ production can be a better test for the perturbative QCD than double $J/\psi$ production.

It worth to mention that the hadronic  $J/\psi$ pair production can be researched at low energies in fixed 
target experiments,  where quark-antiquark interactions mostly contribute 
to this process~\cite{Badier:1982ae,Badier:1985ri}\footnote{the simultaneous production of two $J/\psi$-mesons was first 
observed in 1982 by the NA3 collaboration in the multi-muon events in pion-platinum interactions at 150 and 280 GeV and
later at 400 GeV in proton-platinum collisions. These data is fairly described within the CS model~\cite{Kartvelishvili:1984ur}.}, 
as well as at high energies of LHC collider, where the gluon interaction is dominant.

Contrary to this, the double $\Upsilon$ production can be researched only at the LHC due to the small 
cross section value (which is approximately  $10^{3}$ times smaller than for the $J/\psi$ pair production), as well as the 
large invariant mass needed to produce an $\Upsilon$ pair.
It could be expected, that the gluon fusion gives the main contribution into the process at the LHC. 

In this work we are mainly interested in the low invariant mass region as it is most accessible
in the LHCb experiment, which specializes in heavy quarks studies. We assume, that the 
next leading order corrections are small for the LHCb experiment conditions.  
It should be noted here that  at large invariant masses the high order corrections change the cross section behaviour and it becomes a constant, as well for the double $J/\psi$-production~\cite{Kiselev:1988ww},
while  within the LO of perturbative QCD the cross section for these processes decreases with gluonic energy as $\sim 1/s^2$, as it is shown in~\cite{Humpert:1983yj}.

ATLAS is another LHC experiment, where the $\Upsilon$ pair production can be researched. Despite the
$p_T$ cuts applied for  muons from  $\Upsilon$ decays, ATLAS can reconstruct $\Upsilon$-mesons without
cuts on their $p_T$~\cite{Aad:2011xv} due to the large energy release in the leptonic $\Upsilon$ decay. The efficiency and the total number of reconstructed events in the low-$p_T$ region is not expected to exceed the LHCb values, however the cross section of double $\Upsilon$ production in the ATLAS experiment is predicted to be larger than in LHCb.

It is worth to note that in the leading 
order of $\alpha_s$ the quarkonia pair production obeys selection rules, which are similar to those
for quarkonia decays. Selection rules in the charmonia decays are confirmed by measurements of $J/\psi$ and $\chi_c$ meson widths.
In the production case two initial gluons in a color-singlet state are C-even. This is why the production
of $\Upsilon$-, $\eta_b$- or $\chi_b$-meson pairs is allowed, while the combined production of two
particles having a different C-parity (such as $\Upsilon+\eta_b$ and $\Upsilon+\chi_b$) is prohibited.

These selection rules put restrictions on the feed-down from the higher states. For instance the feed-down from the
$\Upsilon+\chi_{b}$ production is expected to be less than from the $\chi_{b}+\chi_{b}$ one, because the  production of the former final state is forbidden in the leading order of pQCD. Meanwhile  the feed-down from the $\Upsilon+\Upsilon'$ channel can be quite large. 

We expect that the color-octet mechanism negligibly contributes to the double $\Upsilon$ production. 
Indeed, the color-octet contribution to the $\Upsilon$ production should be essentially smaller than 
the color-octet contribution to the $J/\psi$ production, because the color-octet matrix element is suppressed by the second power of relative velocity of the quarks inside the meson. But even for the doubly $J/\psi$ production at LHCb the color-octet should not be taken into account\footnote{As it is shown in \cite{Qiao:2009kg,Ko:2010xy}
the octet contribution becomes significant only at transverse momentum $p_{T} \gtrsim 15$ GeV, which corresponds to a big invariant mass of the $J/\psi$ pair.}.

The $\Upsilon$-meson pair production will be investigated in the decay mode $\Upsilon\to \mu^+ \mu^-$ for the both quarkonia: $2 \Upsilon \to \mu^+ \mu^- \mu^+ \mu^- $. The region of invariant masses of four muons near the 
threshold of the $\Upsilon$-pair production can be interesting due to the opportunity for two diquarks 
$[bb]_{\bar 3}+\left[\bar b \bar b \right]_{3}$ to form a bound state --- a tetraquark, which could decay to 
$\Upsilon +\ \mu^+ \mu^-\to \mu^+ \mu^- \mu^+ \mu^-$.
An attraction between $\bar{3}$ and $3$ color states does not exclude such a possibility, especially since
similar exotic states like the $Y(3940)$ resonance, decaying to $J/\psi\omega$~\cite{Drenska:2010kg}, and 
the $X(4140)$ resonance, decaying to $J/\psi\phi$~\cite{Wick:2010xv,Liu:2010hf}, have recently been observed.
It should be mentioned that according to the LHCb experimental data, there is no evidence of the $4c$-tetraquark, 
decaying into two $J/\psi$ mesons. However $[bb]_{\bar 3}$ is essentially smaller than $[cc]_{\bar 3}$ 
and therefore the model of two attracting doubly heavy diquark can be more applicable for the $4b$-tetraquark.

The second section of our article is devoted to the non-resonant production of $\Upsilon$-meson pairs in 
the gluon-gluon interaction.
In the third section the cross section of this process  at the LHC energy of $8$ TeV  is calculated
taking different experimental restrictions into account. 
The fourth section the production of $4b$-tetraquark at ther LHCb experiment is discussed.

\section{Double $\Upsilon$ production}

The standard method to describe the double $\Upsilon$ production is based on following assumptions:
\begin{enumerate}
\item the LO QCD can be used to calculate a matrix element of four heavy quark production;
\item $b$ and $\bar b$ quarks in $\Upsilon$ are in a color singlet state;
\item the internal motion of quark in the meson do not essentially influence the cross section value 
and the distribution shapes ($\delta$-approximation);
\item the both $\Upsilon$ mesons are produced in a single gluon-gluon interaction (a single parton scattering, SPS). 
\end{enumerate}

In the leading order of perturbative QCD there are 31 Feynman diagrams describing  a color-singlet 
$\Upsilon$ pairs production in the gluonic interaction.

We do not consider the contribution of quark-antiquark interaction, which is negligibly small at LHC energies.

The hadronization of $b\bar b$-pair into the final $\Upsilon$-meson within the $\delta$-approximation is accounted for by the wave function of this particle at origin:
\begin{eqnarray}
\left.\psi_{\Upsilon(1S)}(r)\right|_{r=0} & = & 0.635~\mathrm{GeV^{3/2}},
\nonumber \\
\left.\psi_{\Upsilon(2S)}(r)\right|_{r=0} & = & 0.455~\mathrm{GeV^{3/2}},
\label{eq:psiCC0}
\\
\left.\psi_{\Upsilon(3S)}(r)\right|_{r=0} & = & 0.400~\mathrm{GeV^{3/2}}.
\nonumber
\end{eqnarray}
These are only nonperturbative parameters in the matrix element of $gg\to2\Upsilon$ process. The values~(\ref{eq:psiCC0}) are
extracted from the leptonic widths of $\Upsilon$-mesons, neglecting QCD corrections, as we do not take 
these corrections into account in our matrix element.

As every $\alpha_s^4$ calculation, our calculation of $\Upsilon$ pair production is affected by the scale uncertainty. This one is of the same order of magnitude, as the $b$-quark mass uncertainty. Unlike the scale choice, the choice of the $b$-quark mass also affects on a relative yield of different bottomonia pairs. 

The conventional $b$ mass value in the matrix element equals a half mass of produced bottomonium.  
Alternatively, one can choose one and the same $m_b$ value for all bottomonia pairs.
We present the results obtained for the both these cases.
For the second one we use $m_b =4.73~\mathrm{GeV}$ in the matrix element while the mass of the particular $\Upsilon$-meson is taken in the phase space. 

To estimate the feed-down from higher excitations  we use the PDG~\cite{Beringer:1900zz} branching fractions for $\Upsilon(2S) \to \Upsilon(1S) + X$ and $\Upsilon(3S) \to \Upsilon(1S,2S) + X$ decays.
The impact of feed-down on the cross section values is less than the uncertainties owing to the scale and the $m_b$ choices. Nevertheless, it should be taken into account, because it effects on the relative yields of different bottomonia pairs. The  $\Upsilon(1S)\Upsilon(1S)$ pair production is most influenced by feed-down,  where it leads to the 30\% increase of the yield. 

The  transverse momenta of the initial gluons are taken into account within the  Pythia machinery.

In Tab.~\ref{tbl:totalCS} we present the double  bottomonium  production cross sections (without kinematical cuts) estimated for the different parameters values. All these predictions were done for the 8~TeV pp collisions using  the CTEQ6LL \textit{p.d.f.} set~\cite{Pumplin:2002vw} at the scale equal to an averaged transverse mass of quarkonia.
The $\alpha_s$ in the matrix elements of hard subprocess is also taken at this scale.

\begin{figure}
\includegraphics[width=8cm]{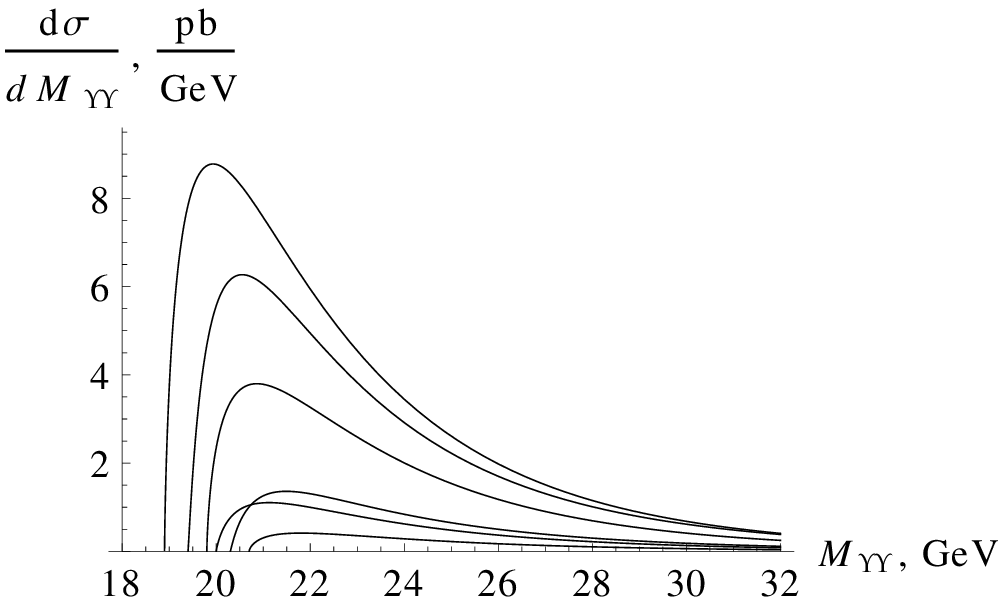} 
\includegraphics[width=8cm]{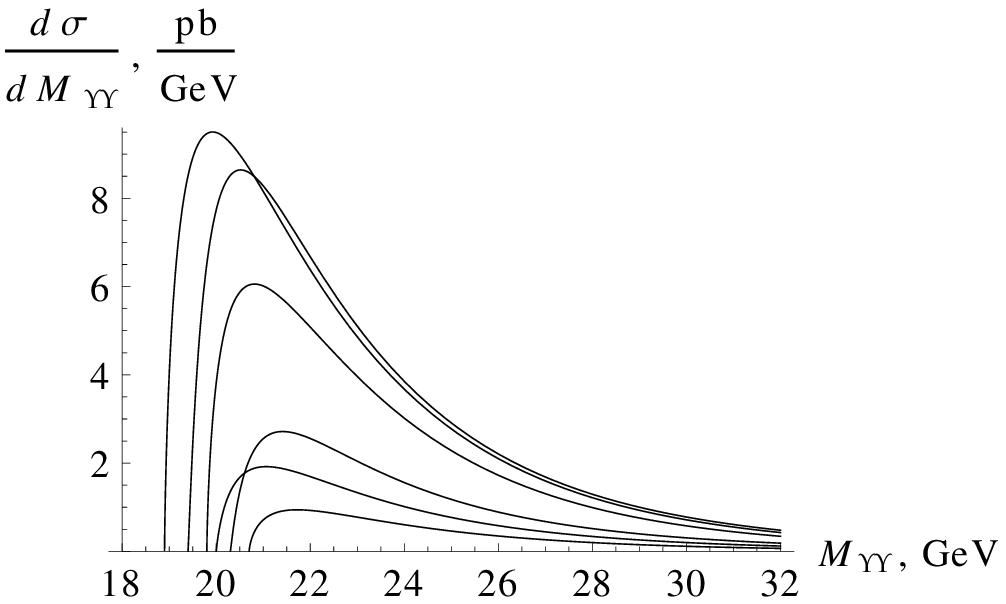} 
\caption{The distributions over the invariant mass of  $\Upsilon$-meson pairs. A half of the meson mass is used for the $m_b$ 
in the left plot and fixed value of $m_b=4.73~\mathrm{GeV}$ is used in the right. From left to right curves correspond to the $\Upsilon(1S)\Upsilon(1S)$, $\Upsilon(1S)\Upsilon(2S)$, $\Upsilon(1S)\Upsilon(3S)$, $\Upsilon(2S)\Upsilon(2S)$,
$\Upsilon(2S)\Upsilon(3S)$ and $\Upsilon(3S)\Upsilon(3S)$ production.
\label{pic:invmassnocut}}
\end{figure}

The corresponding cross section distributions on the invariant mass are shown in Fig.~\ref{pic:invmassnocut}.
The distribution shapes  do not depend essentially on the hard scale choice, while the cross section value changes by factor of approximately $1.4$ with the variation of scale by a factor of $2$.

\begin{table}
\begin{tabular}{|c||c|c|c|c|}
\hline 
\multirow{2}{*}{Mode}		&	\multicolumn{2}{|c|}{$m_b = m_{\Upsilon}/2$}	
							&	\multicolumn{2}{|c|}{$m_b = 4.73~\mathrm{GeV}$}	\\ 
\cline{2-5}
							& without feed-down		& with feed-down		& without feed-down	&	with feed-down		\\ 
\hline 
$\Upsilon(1S)\Upsilon(1S)$	&	$36~\mathrm{pb}$	&$45~\mathrm{pb}$	&$36~\mathrm{pb}$	&	$48~\mathrm{pb}$	\\ 
\hline 
$\Upsilon(2S)\Upsilon(2S)$	&	$5.3~\mathrm{pb}$	&$6.0~\mathrm{pb}$	&$8.6~\mathrm{pb}$	&	$9.8~\mathrm{pb}$	\\ 
\hline 
$\Upsilon(3S)\Upsilon(3S)$	&	$2.3~\mathrm{pb}$	&$2.3~\mathrm{pb}$	&$4.9~\mathrm{pb}$	&	$4.8~\mathrm{pb}$	\\ 
\hline 
$\Upsilon(1S)\Upsilon(2S)$	&	$27~\mathrm{pb}$	&$33~\mathrm{pb}$	&$35~\mathrm{pb}$	&	$44~\mathrm{pb}$	\\ 
\hline 
$\Upsilon(1S)\Upsilon(3S)$	&	$18~\mathrm{pb}$	&$20~\mathrm{pb}$	&$27~\mathrm{pb}$	&	$31~\mathrm{pb}$	\\ 
\hline 
$\Upsilon(2S)\Upsilon(3S)$	&	$7.0~\mathrm{pb}$	&$7.4~\mathrm{pb}$	&$13~\mathrm{pb}$	&	$14~\mathrm{pb}$	\\ 
\hline 
\end{tabular} 
\caption{The double  bottomonium  production cross sections in 8~TeV $pp$-collisions. No kinematical cuts are imposed. A relative uncertainty of $1.4$ due to the scale choice is assumed.
\label{tbl:totalCS}}
\end{table}

Despite the difference in masses and in feed-down contributions, shapes of distributions over 
the transverse momenta and rapidity of different $\Upsilon$ mesons do not differ significantly. At least the uncertainties caused by ambiguity in scale selection are larger. This is why, in what follows,
we discuss kinematic properties of $\Upsilon(1S)$ pair production only.

Here we consider the $\Upsilon$ pair production at LHCb and ATLAS experiments.
   
The LHCb experiment, being designed for the $b$-physics studies, allows to measure the
$\Upsilon$ pair  production without $p_T$ cutoff. Contrary to this, as a rule, the ATLAS experiment allows to study only relatively large $p_T$. However, namely the $\Upsilon$ production is an exclusion. 
Contrary to the $J/\psi$ meson leptonic decay, the energy release in the $\Upsilon$ decay is large.
This allows ATLAS to measure $\Upsilon$ mesons even at low $p_T$. The production cross section in ATLAS 
conditions appears to be even larger than in the LHCb due to the wider rapidity window ($-2.5<y<2.5$ instead of $2<y<4.5$ for LHCb).

\begin{figure}
\includegraphics[width=12cm]{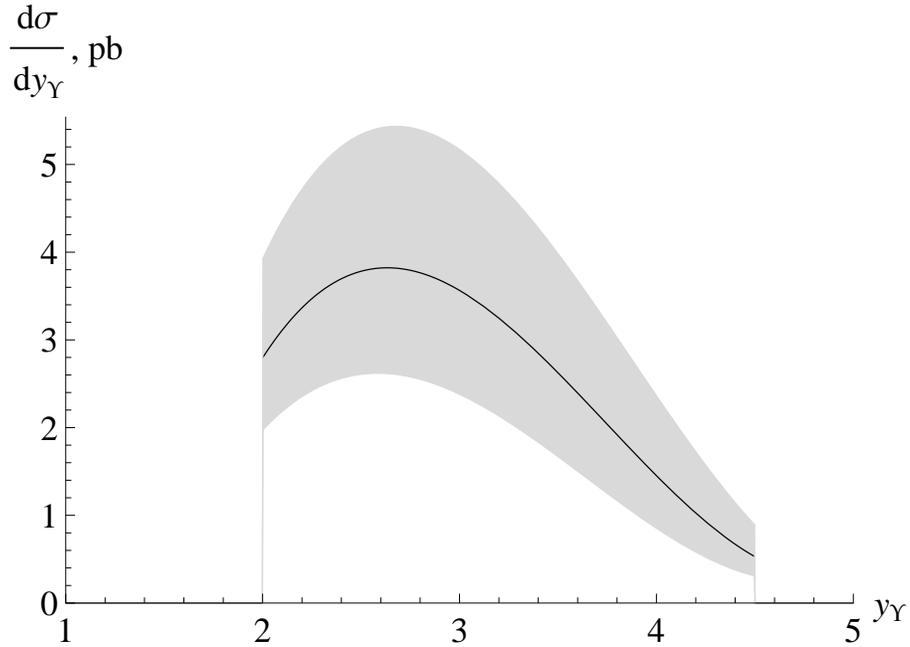} 
\caption{The rapidity distribution of the $\Upsilon$-pair  in the $\Upsilon$-pair
production process with the LHCb rapidity cut. 
\label{pic:yLHCbcut}}
\end{figure}

Within the SPS approach we predict the hump in the rapidity distribution of single $\Upsilon$ meson  in 
the double $\Upsilon$ production at LHCb, as well as the hump in the rapidity distribution of single $J/\psi$ meson   in the double $J/\psi$  production.
This hump is caused by the LHCb cutoff on rapidity, which leads to the strong rapidity correlation (see Fig.~\ref{pic:yLHCbcut}).

\begin{figure}
\includegraphics[width=12cm]{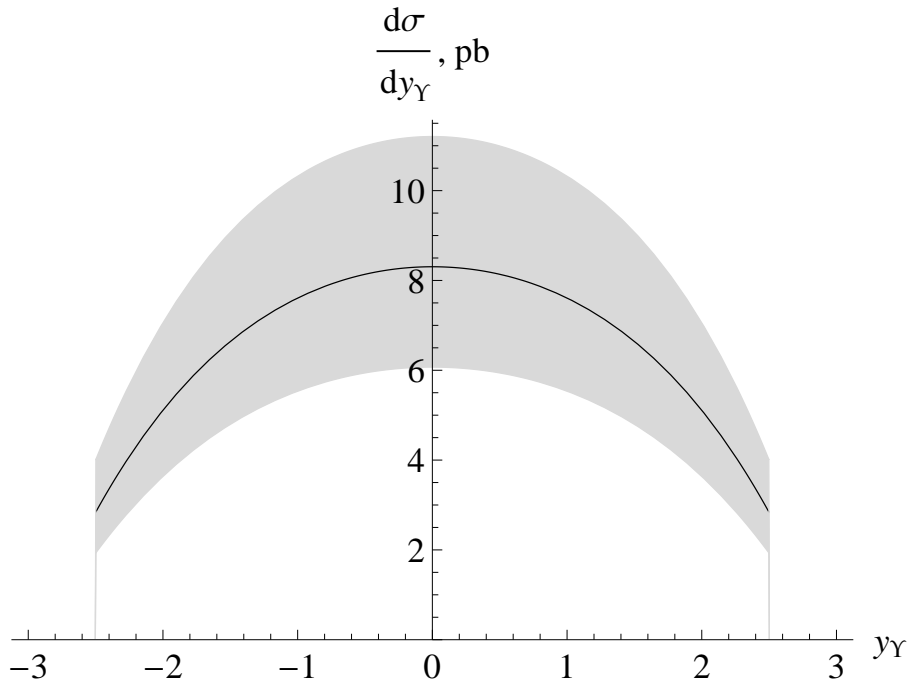} 
\caption{Distribution over the rapidity of single  $\Upsilon$  in the $\Upsilon$-pair
production process at ATLAS.
\label{pic:yATLcut}}
\end{figure}

The rapidity distribution in ATLAS conditions is presented in Fig.~\ref{pic:yATLcut}.

\begin{figure}
\includegraphics[width=12cm]{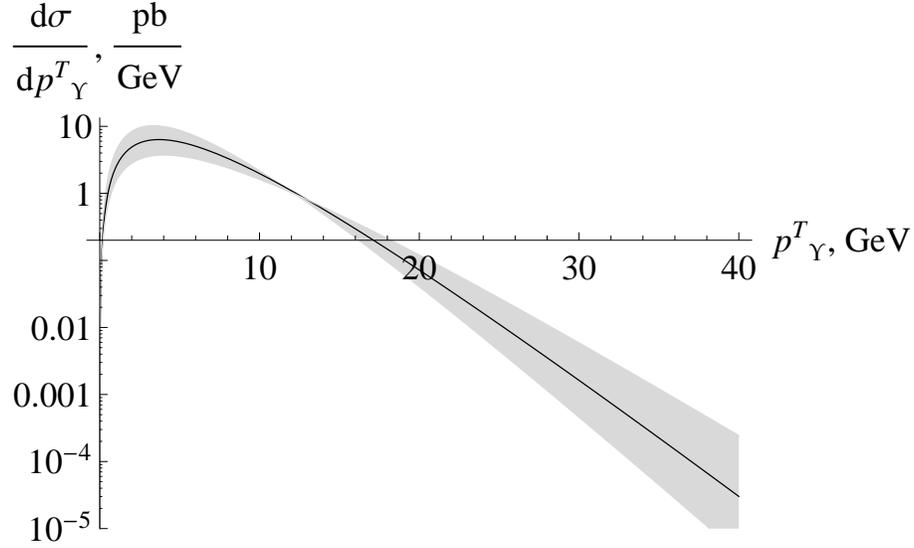} 
\caption{The  $p_T$ distributions  of the single  $\Upsilon$  in the $\Upsilon(1S)\Upsilon(1S)$
production process at LHC depending on the scale choice.
\label{pic:Pt}}
\end{figure}

\begin{figure}
\includegraphics[width=12cm]{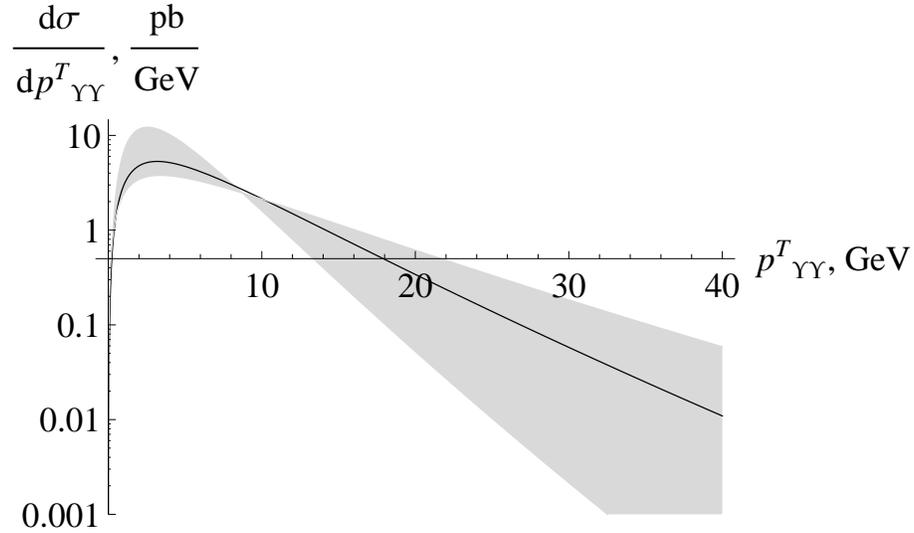} 
\caption{Distribution over the $p_T$ of the $\Upsilon$-pair  in the $\Upsilon(1S)\Upsilon(1S)$
production process at LHC depending on the scale choice.  
\label{pic:PtPair}}
\end{figure}

Despite the essential difference in the kinematical conditions at these facilities  the 
distributions over the $p_T$ of  single $\Upsilon$ in the pair production and over the $p_T$ of  $\Upsilon$ pair have practically the same shapes. This why we perform here these distributions without rapidity cuts  (see Fig.~\ref{pic:Pt} and Fig.~\ref{pic:PtPair}).
 As one can see, the different choices of hard scale value significantly affect 
the slope of  both distributions in the high $p_T$ area. 

\begin{table}
\begin{tabular}{|c||c|c|c|c|}
\hline 
\multirow{2}{*}{Mode}		&	\multicolumn{2}{|c|}{ATLAS}	
							&	\multicolumn{2}{|c|}{LHCb}	\\ 
\cline{2-5}
							&$m_b = m_{\Upsilon}/2$&$m_b = 4.73~\mathrm{GeV}$&$m_b = m_{\Upsilon}/2$&$m_b = 4.73~\mathrm{GeV}$\\ 
\hline 
$\Upsilon(1S)\Upsilon(1S)$	&	$31~\mathrm{pb}$	&$33~\mathrm{pb}$	&$6.2~\mathrm{pb}$	&	$6.6~\mathrm{pb}$	\\ 
\hline 
$\Upsilon(2S)\Upsilon(2S)$	&	$4.2~\mathrm{pb}$	&$6.8~\mathrm{pb}$	&$0.8~\mathrm{pb}$	&	$1.3~\mathrm{pb}$	\\ 
\hline 
$\Upsilon(3S)\Upsilon(3S)$	&	$1.6~\mathrm{pb}$	&$3.4~\mathrm{pb}$	&$0.3~\mathrm{pb}$	&	$0.6~\mathrm{pb}$	\\ 
\hline 
$\Upsilon(1S)\Upsilon(2S)$	&	$23~\mathrm{pb}$	&$30~\mathrm{pb}$	&$4.5~\mathrm{pb}$	&	$6.0~\mathrm{pb}$	\\ 
\hline 
$\Upsilon(1S)\Upsilon(3S)$	&	$14~\mathrm{pb}$	&$21~\mathrm{pb}$	&$2.8~\mathrm{pb}$	&	$4.2~\mathrm{pb}$	\\ 
\hline 
$\Upsilon(2S)\Upsilon(3S)$	&	$5.2~\mathrm{pb}$	&$9.6~\mathrm{pb}$	&$1.0~\mathrm{pb}$	&	$1.9~\mathrm{pb}$	\\ 
\hline 
\end{tabular} 
\caption{The cross sections of doubly bottomonium production in 8~TeV $pp$-collisions. 
ATLAS and LHCb kinematical cuts are imposed.
\label{tab:cutsCS}}
\end{table}

In the LHC conditions a huge density of low-$x$ gluons leads to the increase of multiple gluon-gluon interactions probability within  a single proton-proton collision.
In the DPS approach, which implies a particles production in two independent subprocesses,
the cross section is written down as follows:
\begin{equation} 
\label{eqn:doubleAB}
\sigma^{ A B }_{\rm DPS} = \frac{m}{2} \frac{\sigma^{ A}_{\rm SPS} 
\sigma^{ B}_{\rm SPS}} {\sigma_{\rm eff}}.
\end{equation} 
where the parameter $\sigma_{\rm eff}=14.5~\mathrm{mb}$ was measured in the four jets and three jets plus photon modes by the CDF and D0 detectors \cite{Abe:1997xk,Abazov:2009gc}, and
the parameter $m$  equals $1$ for  identical subprocesses and $2$ for different ones.
As it was shown in~\cite{Kom:2011bd,Baranov:2011ch,Novoselov:2011ff}, 
the DPS mechanism could play an essential role in the double $J/\psi$ production, because
the predictions for the double $J/\psi$ production within SPS and DPS have the same order of magnitude.
For the double $\Upsilon$ production the DPS mechanism contributes only about 10\% of the SPS 
cross section and, therefore, can be neglected   (see~\cite{Novoselov:2011ff}).

\section{The possibility of observation of the $4b$-tetraquark at LHCb}

Within QCD two heavy quarks in the $\bar{3}$ color state attracts each other, forming the $[QQ]_{\bar 3}$ diquark. 
An attraction between $\bar{3}$ and $3$ color states does not exclude a possibility to observe  the heavy tetraquark 
$[QQ]_{\bar 3}[\bar Q \bar Q]_3$. The spectroscopy of such systems can be easily investigated under assumption that 
these diquarks  are almost point-like~\cite{Berezhnoy:2011xn}. 
Within this approach it was predicted that all $4b$-tetraquark states have 
masses below the $\Upsilon+\Upsilon$ threshold. Therefore such states can not contribute to the 
double $\Upsilon$ production.  Nevertheless the $4b$ tetraquark could be searched in the decay mode 
$[bb]_{\bar 3}[\bar b \bar b]_3\to \Upsilon +\Upsilon^*\to \Upsilon \mu^+ \mu^-$. According to~\cite{Berezhnoy:2011xn}
the ground state of $4b$ 
tetraquark splits into tree states due to the spin-spin interaction:
\begin{eqnarray*}
0^{++}: & \qquad & M=18.754~\mathrm{GeV},\qquad M-M_{\mathrm{th}}=-544.~\mathrm{MeV},\\
1^{+-}: & \qquad & M=18.808~\mathrm{GeV},\qquad M-M_{\mathrm{th}}=-490.~\mathrm{MeV},\\
2^{++}: & \qquad & M=18.916~\mathrm{GeV},\qquad M-M_{\mathrm{th}}=-382.~\mathrm{MeV}.\\
\end{eqnarray*}

It worth to note, that at the present time there are no reliable methods to estimate the cross section value of  the $4b$ tetraquark production.

\section{Conclusions}

The observation of $J/\psi$ pair production has opened an interesting discussion about the contributions of SPS and DPS mechanisms  into the production process.  The SPS contribution is estimated in the framework of well known and well tested pQCD approach. The second mechanism, DPS, is not well studied experimentally yet.
 However it allows to understand the large cross sections of the $J/\psi$ plus open charm  production and the double open charm production at LHCb. For the double $J/\psi$ production the SPS and DPS approaches give comparable predictions, which are of the same order with the experimental results.  
 
The situation is different for the double $\Upsilon$ production. 
The both SPS and DPS calculations for it stand on the same basis as for the
double $J/\psi$ production, but for this case the SPS prediction is about an order of magnitude larger  than the DPS one. 
Thus, the study of $\Upsilon$ pair production will allow us to understand, if we describe the SPS production of quarkonia correctly.

The observation of double $J/\psi$-meson 
production~\cite{Aaij:2011yc}
was reported by LHCb soon after the measurement of inclusive $J/\psi$ production~\cite{Aaij:2011jh}.
The measurement of $\Upsilon$-meson inclusive 
production have been recently reported in~\cite{LHCb:2012aa}. It is based on the $25~\mathrm{pb^{-1}}$
integrated luminosity collected in 2010. The integrated luminosity
collected both in 2011 and in 2012  exceeds $1~\mathrm{fb^{-1}}$. This corresponds to more than 40 times lager statistics. Taking into account the estimated here cross section value  and the leptonic branching of   $\Upsilon$-meson ($2.5\%$) one can expect several dozens of events with two $\Upsilon$ mesons in both 2011 and 2012 data.

Authors would like to thank Vanya Belyaev and Alexey Luchinsky for the fruitful discussions. The work was 
 supported by Russian Foundation for Basic Research (grant \#10-02-00061a). The work of Alexey Novoselov was partially supported by the grant of the president of Russian Federation (\#MK-3513.2012.2) and by the non-commercial foundation “Dynasty”.

\bibliography{two_ups}

\begin{thebibliography}{29}
\expandafter\ifx\csname natexlab\endcsname\relax\def\natexlab#1{#1}\fi
\expandafter\ifx\csname bibnamefont\endcsname\relax
  \def\bibnamefont#1{#1}\fi
\expandafter\ifx\csname bibfnamefont\endcsname\relax
  \def\bibfnamefont#1{#1}\fi
\expandafter\ifx\csname citenamefont\endcsname\relax
  \def\citenamefont#1{#1}\fi
\expandafter\ifx\csname url\endcsname\relax
  \def\url#1{\texttt{#1}}\fi
\expandafter\ifx\csname urlprefix\endcsname\relax\def\urlprefix{URL }\fi
\providecommand{\bibinfo}[2]{#2}
\providecommand{\eprint}[2][]{\url{#2}}

\bibitem[{\citenamefont{Aaij et~al.}(2012{\natexlab{a}})}]{Aaij:2011yc}
\bibinfo{author}{\bibfnamefont{R.}~\bibnamefont{Aaij}} \bibnamefont{et~al.}
  (\bibinfo{collaboration}{LHCb}), \bibinfo{journal}{Phys. Lett.}
  \textbf{\bibinfo{volume}{B707}}, \bibinfo{pages}{52}
  (\bibinfo{year}{2012}{\natexlab{a}}), \eprint{1109.0963}.

\bibitem[{\citenamefont{Berezhnoy
  et~al.}(2011{\natexlab{a}})\citenamefont{Berezhnoy, Likhoded, Luchinsky, and
  Novoselov}}]{Berezhnoy:2011xy}
\bibinfo{author}{\bibfnamefont{A.~V.} \bibnamefont{Berezhnoy}},
  \bibinfo{author}{\bibfnamefont{A.~K.} \bibnamefont{Likhoded}},
  \bibinfo{author}{\bibfnamefont{A.~V.} \bibnamefont{Luchinsky}},
  \bibnamefont{and} \bibinfo{author}{\bibfnamefont{A.~A.}
  \bibnamefont{Novoselov}}, \bibinfo{journal}{Phys. Rev.}
  \textbf{\bibinfo{volume}{D84}}, \bibinfo{pages}{094023}
  (\bibinfo{year}{2011}{\natexlab{a}}), \eprint{1101.5881}.

\bibitem[{\citenamefont{Baranov et~al.}(2012)\citenamefont{Baranov, Snigirev,
  Zotov, Szczurek, and Schafer}}]{Baranov:2012re}
\bibinfo{author}{\bibfnamefont{S.}~\bibnamefont{Baranov}},
  \bibinfo{author}{\bibfnamefont{A.}~\bibnamefont{Snigirev}},
  \bibinfo{author}{\bibfnamefont{N.}~\bibnamefont{Zotov}},
  \bibinfo{author}{\bibfnamefont{A.}~\bibnamefont{Szczurek}}, \bibnamefont{and}
  \bibinfo{author}{\bibfnamefont{W.}~\bibnamefont{Schafer}}
  (\bibinfo{year}{2012}), \eprint{1210.1806}.

\bibitem[{\citenamefont{Baranov}(2011)}]{Baranov:2011zz}
\bibinfo{author}{\bibfnamefont{S.}~\bibnamefont{Baranov}},
  \bibinfo{journal}{Phys.Rev.} \textbf{\bibinfo{volume}{D84}},
  \bibinfo{pages}{054012} (\bibinfo{year}{2011}).

\bibitem[{\citenamefont{Martynenko and Trunin}(2012)}]{Martynenko:2012tf}
\bibinfo{author}{\bibfnamefont{A.}~\bibnamefont{Martynenko}} \bibnamefont{and}
  \bibinfo{author}{\bibfnamefont{A.}~\bibnamefont{Trunin}}
  (\bibinfo{year}{2012}), \eprint{1207.3245}.

\bibitem[{\citenamefont{Berezhnoy et~al.}(2012)\citenamefont{Berezhnoy,
  Likhoded, Luchinsky, and Novoselov}}]{Berezhnoy:2012xq}
\bibinfo{author}{\bibfnamefont{A.}~\bibnamefont{Berezhnoy}},
  \bibinfo{author}{\bibfnamefont{A.}~\bibnamefont{Likhoded}},
  \bibinfo{author}{\bibfnamefont{A.}~\bibnamefont{Luchinsky}},
  \bibnamefont{and} \bibinfo{author}{\bibfnamefont{A.}~\bibnamefont{Novoselov}}
  (\bibinfo{year}{2012}), \eprint{1204.1058}.

\bibitem[{\citenamefont{Bodwin et~al.}(1995)\citenamefont{Bodwin, Braaten, and
  Lepage}}]{Bodwin:1994jh}
\bibinfo{author}{\bibfnamefont{G.~T.} \bibnamefont{Bodwin}},
  \bibinfo{author}{\bibfnamefont{E.}~\bibnamefont{Braaten}}, \bibnamefont{and}
  \bibinfo{author}{\bibfnamefont{G.~P.} \bibnamefont{Lepage}},
  \bibinfo{journal}{Phys.Rev.} \textbf{\bibinfo{volume}{D51}},
  \bibinfo{pages}{1125} (\bibinfo{year}{1995}), \eprint{hep-ph/9407339}.

\bibitem[{\citenamefont{Humpert and Mery}(1983)}]{Humpert:1983yj}
\bibinfo{author}{\bibfnamefont{B.}~\bibnamefont{Humpert}} \bibnamefont{and}
  \bibinfo{author}{\bibfnamefont{P.}~\bibnamefont{Mery}}, \bibinfo{journal}{Z.
  Phys.} \textbf{\bibinfo{volume}{C20}}, \bibinfo{pages}{83}
  (\bibinfo{year}{1983}).

\bibitem[{\citenamefont{Aaij et~al.}(2011)}]{Aaij:2011jh}
\bibinfo{author}{\bibfnamefont{R.}~\bibnamefont{Aaij}} \bibnamefont{et~al.}
  (\bibinfo{collaboration}{LHCb}), \bibinfo{journal}{Eur. Phys. J.}
  \textbf{\bibinfo{volume}{C71}}, \bibinfo{pages}{1645} (\bibinfo{year}{2011}),
  \eprint{1103.0423}.

\bibitem[{\citenamefont{Abe et~al.}(1997)}]{Abe:1997xk}
\bibinfo{author}{\bibfnamefont{F.}~\bibnamefont{Abe}} \bibnamefont{et~al.}
  (\bibinfo{collaboration}{CDF}), \bibinfo{journal}{Phys. Rev.}
  \textbf{\bibinfo{volume}{D56}}, \bibinfo{pages}{3811} (\bibinfo{year}{1997}).

\bibitem[{\citenamefont{Abazov et~al.}(2010)}]{Abazov:2009gc}
\bibinfo{author}{\bibfnamefont{V.~M.} \bibnamefont{Abazov}}
  \bibnamefont{et~al.} (\bibinfo{collaboration}{D0}), \bibinfo{journal}{Phys.
  Rev.} \textbf{\bibinfo{volume}{D81}}, \bibinfo{pages}{052012}
  (\bibinfo{year}{2010}), \eprint{0912.5104}.

\bibitem[{\citenamefont{{LHCb Collaboration}}(2012)}]{LHCb-PAPER-2012-003}
\bibinfo{author}{\bibnamefont{{LHCb Collaboration}}} (\bibinfo{year}{2012}),
  \bibinfo{note}{{LHCb-PAPER-2012-003}}.

\bibitem[{\citenamefont{Badier et~al.}(1982)}]{Badier:1982ae}
\bibinfo{author}{\bibfnamefont{J.}~\bibnamefont{Badier}} \bibnamefont{et~al.}
  (\bibinfo{collaboration}{NA3 Collaboration}), \bibinfo{journal}{Phys.Lett.}
  \textbf{\bibinfo{volume}{B114}}, \bibinfo{pages}{457} (\bibinfo{year}{1982}).

\bibitem[{\citenamefont{Badier et~al.}(1985)}]{Badier:1985ri}
\bibinfo{author}{\bibfnamefont{J.}~\bibnamefont{Badier}} \bibnamefont{et~al.}
  (\bibinfo{collaboration}{NA3 Collaboration}), \bibinfo{journal}{Phys.Lett.}
  \textbf{\bibinfo{volume}{B158}}, \bibinfo{pages}{85} (\bibinfo{year}{1985}).

\bibitem[{\citenamefont{Kartvelishvili and
  Esakiya}(1983)}]{Kartvelishvili:1984ur}
\bibinfo{author}{\bibfnamefont{V.~G.} \bibnamefont{Kartvelishvili}}
  \bibnamefont{and} \bibinfo{author}{\bibfnamefont{S.~M.}
  \bibnamefont{Esakiya}}, \bibinfo{journal}{Yad. Fiz.}
  \textbf{\bibinfo{volume}{38}}, \bibinfo{pages}{722} (\bibinfo{year}{1983}).

\bibitem[{\citenamefont{Kiselev et~al.}(1989)\citenamefont{Kiselev, Likhoded,
  Slabospitsky, and Tkabladze}}]{Kiselev:1988ww}
\bibinfo{author}{\bibfnamefont{V.}~\bibnamefont{Kiselev}},
  \bibinfo{author}{\bibfnamefont{A.}~\bibnamefont{Likhoded}},
  \bibinfo{author}{\bibfnamefont{S.}~\bibnamefont{Slabospitsky}},
  \bibnamefont{and}
  \bibinfo{author}{\bibfnamefont{A.}~\bibnamefont{Tkabladze}},
  \bibinfo{journal}{Sov.J.Nucl.Phys.} \textbf{\bibinfo{volume}{49}},
  \bibinfo{pages}{682} (\bibinfo{year}{1989}).

\bibitem[{\citenamefont{Aad et~al.}(2011)}]{Aad:2011xv}
\bibinfo{author}{\bibfnamefont{G.}~\bibnamefont{Aad}} \bibnamefont{et~al.}
  (\bibinfo{collaboration}{ATLAS Collaboration}), \bibinfo{journal}{Phys.Lett.}
  \textbf{\bibinfo{volume}{B705}}, \bibinfo{pages}{9} (\bibinfo{year}{2011}),
  \eprint{1106.5325}.

\bibitem[{\citenamefont{Qiao et~al.}(2010)\citenamefont{Qiao, Sun, and
  Sun}}]{Qiao:2009kg}
\bibinfo{author}{\bibfnamefont{C.-F.} \bibnamefont{Qiao}},
  \bibinfo{author}{\bibfnamefont{L.-P.} \bibnamefont{Sun}}, \bibnamefont{and}
  \bibinfo{author}{\bibfnamefont{P.}~\bibnamefont{Sun}},
  \bibinfo{journal}{J.Phys.G} \textbf{\bibinfo{volume}{G37}},
  \bibinfo{pages}{075019} (\bibinfo{year}{2010}), \eprint{0903.0954}.

\bibitem[{\citenamefont{Ko et~al.}(2011)\citenamefont{Ko, Yu, and
  Lee}}]{Ko:2010xy}
\bibinfo{author}{\bibfnamefont{P.}~\bibnamefont{Ko}},
  \bibinfo{author}{\bibfnamefont{C.}~\bibnamefont{Yu}}, \bibnamefont{and}
  \bibinfo{author}{\bibfnamefont{J.}~\bibnamefont{Lee}},
  \bibinfo{journal}{JHEP} \textbf{\bibinfo{volume}{01}}, \bibinfo{pages}{070}
  (\bibinfo{year}{2011}), \eprint{1007.3095}.

\bibitem[{\citenamefont{Drenska et~al.}(2010)\citenamefont{Drenska, Faccini,
  Piccinini, Polosa, Renga et~al.}}]{Drenska:2010kg}
\bibinfo{author}{\bibfnamefont{N.}~\bibnamefont{Drenska}},
  \bibinfo{author}{\bibfnamefont{R.}~\bibnamefont{Faccini}},
  \bibinfo{author}{\bibfnamefont{F.}~\bibnamefont{Piccinini}},
  \bibinfo{author}{\bibfnamefont{A.}~\bibnamefont{Polosa}},
  \bibinfo{author}{\bibfnamefont{F.}~\bibnamefont{Renga}},
  \bibnamefont{et~al.}, \bibinfo{journal}{Riv.Nuovo Cim.}
  \textbf{\bibinfo{volume}{033}}, \bibinfo{pages}{633} (\bibinfo{year}{2010}),
  \eprint{1006.2741}.

\bibitem[{\citenamefont{Wick}(2009)}]{Wick:2010xv}
\bibinfo{author}{\bibfnamefont{F.}~\bibnamefont{Wick}}
  (\bibinfo{collaboration}{CDF Collaboration}), \bibinfo{journal}{PoS}
  \textbf{\bibinfo{volume}{EPS-HEP2009}}, \bibinfo{pages}{085}
  (\bibinfo{year}{2009}), \eprint{1011.0616}.

\bibitem[{\citenamefont{Liu et~al.}(2011)\citenamefont{Liu, Luo, and
  Zhu}}]{Liu:2010hf}
\bibinfo{author}{\bibfnamefont{X.}~\bibnamefont{Liu}},
  \bibinfo{author}{\bibfnamefont{Z.-G.} \bibnamefont{Luo}}, \bibnamefont{and}
  \bibinfo{author}{\bibfnamefont{S.-L.} \bibnamefont{Zhu}},
  \bibinfo{journal}{Phys.Lett.} \textbf{\bibinfo{volume}{B699}},
  \bibinfo{pages}{341} (\bibinfo{year}{2011}), \eprint{1011.1045}.

\bibitem[{\citenamefont{Beringer et~al.}(2012)}]{Beringer:1900zz}
\bibinfo{author}{\bibfnamefont{J.}~\bibnamefont{Beringer}} \bibnamefont{et~al.}
  (\bibinfo{collaboration}{Particle Data Group}), \bibinfo{journal}{Phys.Rev.}
  \textbf{\bibinfo{volume}{D86}}, \bibinfo{pages}{010001}
  (\bibinfo{year}{2012}).

\bibitem[{\citenamefont{Pumplin et~al.}(2002)}]{Pumplin:2002vw}
\bibinfo{author}{\bibfnamefont{J.}~\bibnamefont{Pumplin}} \bibnamefont{et~al.},
  \bibinfo{journal}{JHEP} \textbf{\bibinfo{volume}{07}}, \bibinfo{pages}{012}
  (\bibinfo{year}{2002}), \eprint{hep-ph/0201195}.

\bibitem[{\citenamefont{Kom et~al.}(2011)\citenamefont{Kom, Kulesza, and
  Stirling}}]{Kom:2011bd}
\bibinfo{author}{\bibfnamefont{C.~H.} \bibnamefont{Kom}},
  \bibinfo{author}{\bibfnamefont{A.}~\bibnamefont{Kulesza}}, \bibnamefont{and}
  \bibinfo{author}{\bibfnamefont{W.~J.} \bibnamefont{Stirling}},
  \bibinfo{journal}{Phys. Rev. Lett.} \textbf{\bibinfo{volume}{107}},
  \bibinfo{pages}{082002} (\bibinfo{year}{2011}), \eprint{1105.4186}.

\bibitem[{\citenamefont{Baranov et~al.}(2011)\citenamefont{Baranov, Snigirev,
  and Zotov}}]{Baranov:2011ch}
\bibinfo{author}{\bibfnamefont{S.}~\bibnamefont{Baranov}},
  \bibinfo{author}{\bibfnamefont{A.}~\bibnamefont{Snigirev}}, \bibnamefont{and}
  \bibinfo{author}{\bibfnamefont{N.}~\bibnamefont{Zotov}},
  \bibinfo{journal}{Phys.Lett.} \textbf{\bibinfo{volume}{B705}},
  \bibinfo{pages}{116} (\bibinfo{year}{2011}), \eprint{1105.6276}.

\bibitem[{\citenamefont{Novoselov}(2011)}]{Novoselov:2011ff}
\bibinfo{author}{\bibfnamefont{A.}~\bibnamefont{Novoselov}}
  (\bibinfo{year}{2011}), \eprint{1106.2184}.

\bibitem[{\citenamefont{Berezhnoy
  et~al.}(2011{\natexlab{b}})\citenamefont{Berezhnoy, Luchinsky, and
  Novoselov}}]{Berezhnoy:2011xn}
\bibinfo{author}{\bibfnamefont{A.}~\bibnamefont{Berezhnoy}},
  \bibinfo{author}{\bibfnamefont{A.}~\bibnamefont{Luchinsky}},
  \bibnamefont{and} \bibinfo{author}{\bibfnamefont{A.}~\bibnamefont{Novoselov}}
  (\bibinfo{year}{2011}{\natexlab{b}}), \eprint{1111.1867}.

\bibitem[{\citenamefont{Aaij et~al.}(2012{\natexlab{b}})}]{LHCb:2012aa}
\bibinfo{author}{\bibfnamefont{R.}~\bibnamefont{Aaij}} \bibnamefont{et~al.}
  (\bibinfo{collaboration}{LHCb Collaboration}), \bibinfo{journal}{Eur.Phys.J.}
  \textbf{\bibinfo{volume}{C72}}, \bibinfo{pages}{2025}
  (\bibinfo{year}{2012}{\natexlab{b}}), \eprint{1202.6579}.

\end{thebibliography}
\end{document}